**Peptides of *H. sapiens* and *P. falciparum* that are predicted to bind strongly to HLA-A*24:02 and homologous to a SARS-CoV-2 peptide**

Yekbun Adiguzel[a]*

[a] Department of Biophysics, School of Medicine, Altinbas University, Kartaltepe Mah. Incirli Cad. No11, 34147 Bakirkoy, Istanbul, Turkey. E-mail: yekbun.adiguzel@altinbas.edu.tr

**Abstract**

*Aim* - This study is looking for a common pathogenicity between SARS-CoV-2 and Plasmodium species, in individuals with certain HLA serotypes.

*Methods* - *1*. Tblastx searches of SARS-CoV-2 are performed by limiting searches to five Plasmodium species that infect humans. *2*. Aligned sequences in the respective organisms' proteomes are searched with blastp. *3*. Binding predictions of the identified SARS-CoV-2 peptide to HLA supertype representatives are performed. *4*. Blastp searches of predicted epitopes that bind strongly to the identified HLA allele are performed by limiting searches to *H. sapiens* and Plasmodium species, separately. *5*. Peptides with minimum 60% identity to the predicted epitopes are found in results. *6*. Peptides among those, which bind strongly to the same HLA allele, are predicted. *7*. Step-*4* is repeated by limiting searches to *H. sapiens*, followed by the remaining steps until step-*7*, for peptides sourced by Plasmodium species after step-*6*.

*Results* - SARS-CoV-2 peptide with single letter amino acid code CFLGYFCTCYFGLFC has the highest identity to *P. vivax*. Its YFCTCYFGLF part is predicted to bind strongly to HLA-A*24:02. Peptides in the human proteome both homologous to YFCTCYFGLF and with a strong binding affinity to HLA-A*24:02 are YYCARRFGLF, YYCHCPFGVF, and YYCQQYFFLF. Such peptides in the Plasmodium species' proteomes are FFYTFYFELF, YFVACLFILF, and YFPTITFHLF. The first one belonging to *P. falciparum* has a homologous peptide (YFYLFSLELF) in the human proteome, which also has a strong binding affinity to the same HLA allele.

*Conclusion* - Immune responses to the identified-peptides with similar sequences and strong binding affinities to HLA-A*24:02 can be related to autoimmune response risk in individuals with HLA-A*24:02 serotypes, upon getting infected with SARS-CoV-2 or *P. falciparum*.

**Keywords:** SARS-CoV-2, Plasmodium, human leukocyte antigen, molecular mimicry, autoimmunity, disease susceptibility

# 1. Introduction

COVID-19 pandemic is an ongoing global crisis with devastating effects. Chloroquine derivatives and hydroxychloroquine were among the treatment options (Perricone et al., 2020). However, they were not deemed significantly effective in decreasing death rates or mechanical ventilation risk (Magagnoli et al., 2020), in addition to serious side effects (Touret and De Lamballerie, 2020). Chloroquine and hydroxychloroquine are registered compounds for the treatment and prevention of malaria and for treatment of autoimmune diseases (Perricone et al., 2020). Accordingly, this study is initiated by searching for a relationship between SARS-CoV-2 and five Plasmodium species that cause malaria in humans (Blanquart and Gascuel, 2011; Déchamps et al., 2010), followed by investigating potential of autoimmune reactions through molecular mimicry. In relation, SARS-CoV-2 and *Plasmodium falciparum* common immunodominant regions were suggested to be explaining low COVID-19 incidence in the "malaria-endemic belt" (Iesa et al., 2020). That study was initiated mainly due to the small number of confirmed cases of COVID-19 at the African region with high malaria-prevalence (*references 4–6 of* (Iesa et al., 2020)). This lower incidence was highlighted in an earlier study as well (*references 1,2 of* (Mehta et al., 2020)). The presence of zoonotic viral genome in *P. vivax* (Charon et al., 2019), which was pointed also by others (De Souza, 2020), was mentioned as a basis of related work (Mehta et al., 2020). Also, one case study reported co-infection with COVID-19 and *P. vivax* in a 10-year-old boy (Kishore et al., 2020). The boy received incomplete primaquine therapy six months before, for the *P. vivax* infection. Authors concluded that it was a relapse of earlier infection, triggered by SARS-CoV-2. Yet, re-infection with *P. vivax* simultaneously with COVID-19 or a relapse in its natural course was not ruled out. However, incidences of malaria and SARS-CoV-2 were found to have a negative correlation (Mehta et al., 2020). The genomic similarity between various strains of SARS-CoV-2 and the matryoshka RNA virus 1 (MaRNAV-1), associated with *P. vivax*, was also investigated (Charon et al., 2019). Accordingly, 15-base pair (bp) alignment with 93.75% identity (e-value 3.1) between SARS-CoV-2 (NC_045512.2) 22487–22502 bp region and 2796–2781 bp region of MaRNAV-1 was observed. The aligned region, which is encoding surface glycoprotein, was suggested to be small to draw any conclusion. Yet, authors stated that a shared immunogenicity cannot be ruled out. Further, their observation was valid in all different strains' data that was used. Cross-species' peptide sharing and high number of matches of amino acids (aa) are expected. Ultimate claim of this work is that such similarities (homologies) carry the potential of leading to inherently related autoimmune pathologies.

The immune response to pathogen proteins is a defence mechanism of the host. However, it can cause adverse effects when pathogen protein has similar sequences to that of humans. That is molecular mimicry, which can lead to autoimmune response. Researchers (Kerkar and Vergani, 2018) supported this notion by findings of *de novo* autoimmune hepatitis associated with certain viral infections. Earlier, related work on HIV supported the concept of immune network manipulation and autoimmunity induced by HIV, through mechanisms involving molecular mimicry (Metlas et al., 1994; Metlas et al., 1999a; Metlas et al., 1999b). Associated work pointed at related problems in HIV-1 gp120-based vaccines (Veljkovic et al., 1997). Similarities and pathway analysis of homologous peptides implied viral infection associated immune response pathologies (Carter, 2011). Kanduc and Shoenfeld contributed to the field with several related works (Kanduc and Shoenfeld, 2016; Kanduc and Shoenfeld, 2018; Kanduc and Shoenfeld, 2019) and they considered risks inherent in associated vaccines, investigated vaccines for papilloma virus L1 (HPV L1) proteins of 4 strains, and surface antigen of hepatitis B virus (Kanduc and Shoenfeld, 2016). Cross-reactions of immune responses against HPV L1 proteins with 20 human proteins were deemed feasible, revealing a probable connection with pathophysiological conditions, implying that peptide-based vaccines need to be pathogen-specific. Regarding SARS-CoV-2, researchers found homologous sequences between spike glycoprotein of the virus, and human surfactant and related proteins, indicative of associated pathologies (Kanduc and Shoenfeld, 2020a). In relation, autoimmune reaction risk of vaccines based on whole SARS-CoV-2 antigens was told to have less likelihood of being revealed through tests on primates. This expectation is due to lower peptide sharing between primates and human pathogens, including SARS-CoV-2 (Kanduc and Shoenfeld, 2020b). Researchers suggested "aged mice" as suitable animal model for testing SARS-CoV-2 spike glycoproteins-based vaccines (Kanduc and Shoenfeld, 2020c).

The relationship of COVID-19 with immunity is multifaceted (Atyeo et al., 2020; Kaneko et al., 2020; Kuri-Cervantes et al., 2020; Laing et al., 2020; Lucas et al., 2020; Mathew et al., 2020). COVID-19 patients experiencing serious health threats displayed hallmarks of extrafollicular B cell activation like in autoimmune settings (Woodruff et al., 2020). Autoinflammatory and autoimmune conditions in COVID-19 are also reviewed (Rodríguez et al., 2020). COVID-19 was suggested to be linked to autoimmunity through mechanisms such as molecular mimicry and bystander activation. SARS-CoV-2 was also hypothesised to be triggering stress-induced autoimmunity through molecular mimicry (Cappello et al., 2020). Also, studies on cerebrospinal fluids of COVID-19 patients were indicative of autoimmunity (Lucchese, 2020). Peptide sharing between SARS-CoV-2 proteome and human brainstem respiratory pacemaker proteins (Lucchese and Flöel, 2020a); human heat shock proteins 90 and 60 (Lucchese and Flöel, 2020b); Odorant Receptor 7D4, Poly ADP-Ribose Polymerase Family Member 9, and Solute Carrier Family 12 Member 6 proteins of human (Angileri et al., 2020) were reported. Strong immune cross-reactions with SARS-CoV-2 spike protein antibody (Vojdani and Kharrazian, 2020) can infer autoimmunity risk in susceptible individuals, through molecular mimicry (Rodríguez et al., 2020). More than one-third of the immunogenic peptides of SARS-CoV-2 have homology with proteins that are of importance for adaptive immune system (Lyons-Weiler, 2020). Also, various disorders associated with possible autoimmunity against human peptides homologous to immunogenic SARS-CoV-2 peptides were pointed at, and examination of patients' sera for autoantibodies against those human peptides were suggested (Kanduc, 2020).

Certain major histocompatibility complex (MHC) allele types can cause risk in certain diseases. For instance, patients expressing human leukocyte antigen (HLA)-DRB1*04:01 are suffering from severe form of rheumatoid arthritis more, and in relation, HLA-DRB1*04:01 has a 5mer, which is shared with *E. coli* heat shock protein (Rojas et al., 2018). This similarity can bear risk in rheumatoid arthritis patients with HLA-DRB1*04:01, when exposed to Enterobacteriaceae. Besides, pathogen and human proteins' peptide similarities are present in systemic lupus erythematosus (James and Harley, 1992), systemic sclerosis (Lunardi et al., 2000), and primary biliary cholangitis (Fussey et al., 1990). Regarding SARS-CoV-2, conserved viral proteome regions with high MHC allele-binding affinities are not found under positive or negative selective pressure (Nguyen et al., 2020). Nevertheless, different studies (e.g., (Cappello, 2020a; Cappello, 2020b; Sedaghat and Karimi, 2020), in addition to those mentioned) suggested related autoimmune reaction risk. Also, association of several alleles through exclusive routes can prevent sorting the risk posed by individual allele types in population-based studies. Additionally, HLA-B*46:01 (Lin et al., 2003; Alicia, 2020) and HLA-B*07:03 (Ng et al., 2004; Alicia, 2020) were found to be leading to susceptibility to the associated disease, SARS.

In Plasmodium infections, pathogen-infected cell-detection by cytotoxic CD8(+) T cells of adaptive immune system was thought to be protective through MHC class I molecules in the liver-stage malaria, but not in the blood stage malaria because erythrocytes do not express MHC class I molecules (Rivera-Correa and Rodriguez, 2020). Finding that *P. vivax*-infected reticulocytes keep expressing MHC class I molecules changed this concept (Junqueira et al., 2018). Regarding molecular mimicry, that between histamine-releasing factor of human and translationally controlled tumour protein of *P. falciparum* (MacDonald et al., 2001); vitronectin and erythrocyte membrane protein 1 of *P. falciparum* (Ludin et al., 2011); and ankyrin, spectrin, and actin proteins of erythrocytes and *P. vivax* proteins (Mourão et al., 2018), were mentioned (Mourão et al., 2020).

## 2. Methods

*2.1 Blast search*

Tblastx (Altschul et al., 1997) at NCBI (NCBI Resource Coordinators, 2017) is used to compare SARS-CoV-2 (NC_045512.2) and five Plasmodium species that infect humans (*P.*

*falciparum*, taxid:5833; *P. malariae*, taxid:5858; *P. vivax*, taxid:5855; *P. ovale*, taxid:36330; *P. knowlesi*, taxid:5850). Separate searches are performed by limiting searches to individual Plasmodium species. Default algorithm parameters are used, except for the number of maximum target sequences. Accordingly, the highest number of allowed target sequences was used. It was 20000 in the first searches and diminished to 5000 later, by the server, for all blast searches.

Aligned sequences with the highest identity in the tblastx outputs are used as inputs of blastp search at NCBI, performed to check the presence of those peptides in the respective organisms' proteomes. Blastp searches are also performed using default parameters, except for increasing the number of maximum target sequences, as mentioned. Sequences are displayed as single letter amino acid (aa) codes.

*2.2 MHC class I-binding*

SARS-CoV-2 peptide identified in the previous step is used as input for predicting its binding affinities to MHC class I (*MHC class I genes are HLA-A, -B, and -C genes*) (Nguyen et al., 2020) proteins. NetMHC 4.0 (Andreatta and Nielsen, 2016; Nielsen et al., 2003), NetMHCpan 4.1 (Reynisson et al., 2020), PickPocket 1.1 (Zhang et al., 2009), and NetMHCcons 1.1 (Karosiene et al., 2012) are the tools used for predictions. Predictions for binding affinities to 12 HLA supertype representatives (HLA-A*01:01 [A1], HLA-A*02:01 [A2], HLA-A*03:01 [A3], HLA-A*24:02 [A24], HLA-A*26:01 [A26], HLA-B*07:02 [B7], HLA-B*08:01 [B8], HLA-B*27:05 [B27], HLA-B*39:01 [B39], HLA-B*40:01 [B44], HLA-B*58:01 [B58], HLA-B*15:01 [B62]) are performed. Fasta format is used as the input format, to be able to choose different peptide lengths, into which input sequence is chopped. Then, all peptide lengths are selected, and default parameters are used for the rest. In predictions with NetMHCpan, BA option is also used for binding affinities to be included in the outputs. Only strong binders (SBs) are considered in the analysis of outputs. In PickPocket, 1-log50k(aff) is evaluated manually, by considering those ≥ 0.5 as strong binder (SB). 1-log50k(aff) is calculated by taking the difference of logarithm (with base 50000) of the affinity (in nM) from 1. Possible difference in the outcome due to using a peptide as input is checked using protein sequence containing the peptide sequence, as input.

NetCTLpan 1.1 is used to predict cytotoxic T lymphocyte (CTL) epitopes (Larsen et al., 2007; Stranzl et al., 2020). It is used similarly to predict epitopes that bind with 12 HLA supertypes. Default parameters are used as before, but by performing predictions for different peptide lengths separately, as required by the tool. However, change in the outcome due to using a short sequence as input is checked again, by using protein sequence containing the peptide sequence, as input. The aim there is to adjust C-term cleavage weight in the epitope prediction, in case that a significant difference is observed.

Peptides identified as SBs are considered based on the following criteria: If predictions by at least three of the tools estimate such a binding affinity to the same HLA allele, in addition to epitope prediction by NetCTLpan. Those peptides are termed as SB-peptides. Finally, information at the websites of predictors is suggested for further details on the tools.

*2.3 Blastp search of SB-peptide and specific HLA allele-binding of SB-peptide homologues*

*2.3.1 Blastp search of SB-peptide and identifying SB-peptide homologues*

Peptide sequence used as input in section 2.2 is searched with blastp, as described in section 2.1, by limiting search to *H. sapiens* (taxid:9606). SB-peptides of the specific HLA allele are searched in blastp results, as follows: Sequences within the results of blastp search are deemed as homologous if they align with query sequence with more than 60% match. Simultaneously, they are expected to contain whole SB-peptide sequence within the query, while there would be no gap in the aligned subject sequence. Analysis is done with the Excel Program of MS Office.

*2.3.2 Specific HLA allele-binding predictions of SB-peptide homologues*

After the identification of homologous sequences of SB-peptides, their binding to the respective HLA serotype is predicted as described in section 2.2, but by choosing the same length as that of SB-peptide, and the specific HLA allele of SB-peptide. The remaining is the same as described in section 2.2. The prediction with the protein sequence containing SB-peptide homologue is not performed here.

Homologous peptides of SB-peptide are searched in the Plasmodium species' proteomes as well. For that, blastp searches are performed as described in section 2.1, by limiting searches to Plasmodium species. Homologous sequences of SB-peptides within the alignment results are identified if they had more than 60% match, while there is no gap in the aligned subject sequence. Following that, the binding affinities of resulting peptide sequences to the HLA allele of SB-peptide are predicted, as described above. The resulting peptide with strong binding affinity to the same HLA allele as SB-peptide is termed SB'-peptide. Blastp search of SB'-peptide is performed eventually, by limiting search to *H. sapiens*. This step is followed by identifying homologous sequences of SB'-peptide that have more than 60% match, while there is no gap in the aligned subject sequence. Finally, the binding affinity of the resulting peptide sequence to the same HLA allele as SB'-peptide (and as SB-peptide) is predicted again, like in section 2.3.2.

*2.4 Visualisation*

TCRpMHCmodels (Jensen et al., 2019) is used to predict the three-dimensional structure of T cell receptor (TCR) alpha and beta chains that are complexed with SB-peptide and MHC class I. Sequences are retrieved from the Protein Data Bank (Berman et al., 2000). TCR alpha and beta chains' sequences are retrieved from the data of H27-14 TCR that is specific for HLA-A24-Nef134-10 (PDBid: 3VXQ; (Shimizu et al., 2013)). MHC's sequence information is retrieved from the data titled "Crystal structure of peptide-HLA-A24 bound to S19-2 V-delta/V-beta TCR" (PDBid: 5XOV; (Shi et al., 2017)).

Seq2Logo 2.0 is a sequence logo generator (Thomsen and Nielsen, 2012). It is used for generating sequence profiles of peptides that are SB to the same HLA allele as SB-peptide and have at least 60% similar sequences with SB-peptide. For logo generation, default parameters are used, other than that for pseudo-count correction (weight on prior) for low counts (see https://services.healthtech.dtu.dk/service.php?Seq2Logo-2.0 for further explanation). Weight on prior option among the adjustable parameters is set to 1 instead of setting to 0, despite expectedly low number of observations. This choice ends up in a logo that appears comparable to the sequence logo of the HLA allele received from MHCMotifViewer (Rapin et al., 2008).

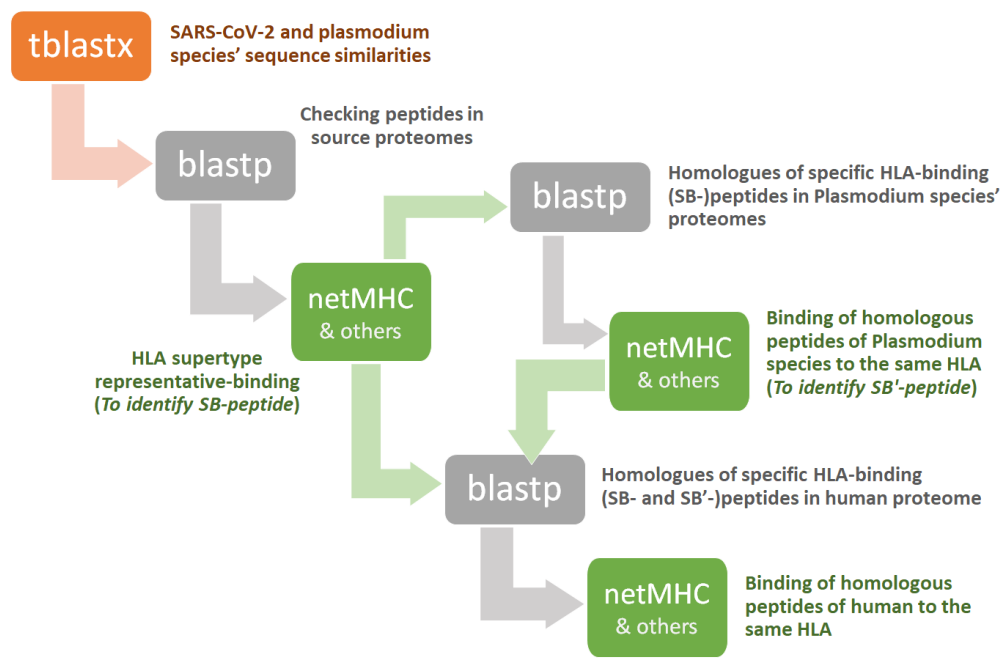

**Fig. 1.** The outline of the process that is followed.

## 3. Results

*3.1 Blast search*

In tblastx (Altschul et al., 1997), query and subject nucleotide sequences are translated in six reading frames, followed by protein blastp comparisons (Wheeler and Bhagwat, 2007). Tblastx retrieved no results for *P. falciparum*. Whereas 1 alignment with the highest 70% identity in case of *P. malariae* (*supplementary file* (*s*)*1*), 12 alignments with the highest 73% identity in case of *P. vivax* (*s2*), 1 alignment with the highest 75% identity in case of *P. ovale* (*s3*), and 3 alignments with the highest 67% identity in case of *P. knowlesi* (*s4*), are observed within alignment results of the others. Aligned sequences' lengths are 10, 15, 8, and 12 aa, respectively. Sequences of those alignments are displayed in Table 1. It can be seen there that *P. knowlesi* sequences VVLWLCCLCWLC and VVLWLCCCCWLC both aligned with the LVLWLLCNCFLC sequence of SARS-CoV-2, with the same maximum identities. Palindromic parts are also present in these aligned query and subject sequences (Table 1, e.g., underlined regions in CNFFNCHYFR, CFLGYFCTCYFGLFC, CFVGYFCTCFVGYFC, FCFCSCFC, LVLWLLCNCFLC, VVLWLCCLCWLC and VVLWLCCCCWLC).

**Table 1.** Aligned SARS-CoV-2 query sequences and Plasmodium subject sequences with the highest identities.

| Query (COVID-19) sequence | Subject (Plasmodium) sequence | Subject name |
| --- | --- | --- |
| - | - | *P. falciparum* |
| seq.1  CNFFNCHYFR | seq.1'  CHYFRCHYFR | *P. malariae* |
| seq.2  CFLGYFCTCYFGLFC | seq.2'  CFVGYFCTCFVGYFC | *P. vivax* |
| seq.3  FCFHKCFC | seq.3'  FCFCSCFC | *P. ovale* |
| seq.4  LVLWLLCNCFLC | seq.4'  VVLWLCCLCWLC | *P. knowlesi* |
|  | seq.4"  VVLWLCCCCWLC |  |

Repetitive regions are present in the sequences displayed in Table 1 (e.g., CHYFR of seq.1', CFVGYFC of seq.2', FC and CFC of seq.3', and WLC of seq.4' and seq.4''). Detection of repetitive regions in comparisons of distant genomes is expected. However, low complexity regions were filtered during tblastx search, as indicated for default algorithm parameters. After analysing results of blastp searches of sequences in Table 1, only CFLGYFCTCYFGLFC sequence is found to be present in translation products of SARS-CoV-2 (*s6* of *s5–13*). Not observing the other sequences in the respective proteomes can be because tblastx compares query and subject after translating nucleotide sequences of each of them in six possible different reading frames. Yet, it can be due to some annotation amendment requirements as well (Böhme et al., 2019).

CFLGYFCTCYFGLFC sequence is not just the query sequence with the highest identity to *P. vivax*. That sequence is present in the alignment results of all respective tblastx searches for SARS-CoV-2 (*s14–17*). Furthermore, it is the most frequently observed query sequence in the alignments (Fig. 2, Table 2, *s18*). When *P. malariae* is selected as the organism in tblastx search of SARS-CoV-2, other alignment results than that with the highest identity contain whole or part of CFLGYFCTCYFGLFC in several queries (*s14, s18*). YFCT repeat is abundant in the alignments among alignment-hits with *P. malariae* genome assembly chromosome 10 (sequence ID LT594498.1) (*s14*). So, subject sequences in the alignments contain repeats, i.e., low complexity regions. It was mentioned that there are 12 alignments with the highest (73%) identity to CFLGYFCTCYFGLFC, within alignment results of tblastx search of SARS-CoV-2, limited to *P. vivax* (*s15*). That tblastx search result has several other alignments with the same sequence, as part or whole of aliged query (i.e., 73 alignments of a total number of 143 alignments, see Table 2) (*s15, s18*). In case of *P. ovale* and *P. knowlesi*, situation is like that of *P. malariae* (*s16, s17*), but the number of alignment-hits with whole or part of CFLGYFCTCYFGLFC is less (Table 2).

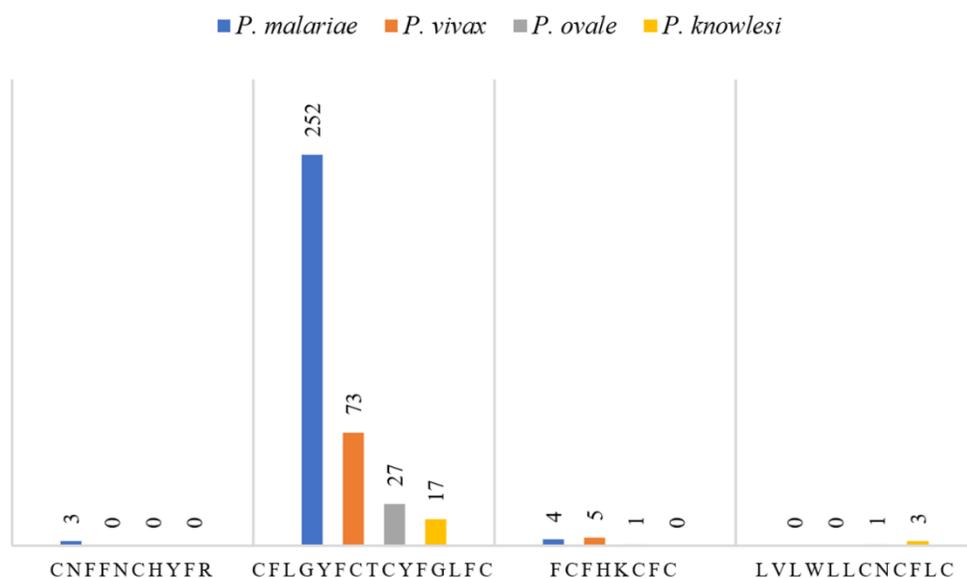

**Fig. 2.** The number of alignments of whole or part of query sequences with the highest percent identities, within alignment results of tblastx searches of SARS-CoV-2, limited to Plasmodium species.

**Table 2.** Alignment frequencies of whole or part of query sequences with the highest percent identities, within alignment results of tblastx searches of SARS-CoV-2, limited to Plasmodium species.•

|                  | *P. malariae* | *P. vivax* | *P. ovale* | *P. knowlesi* |
|------------------|---------------|------------|------------|---------------|
| CNFFNCHYFR       | 3 / 693       | 0 / 143    | 0 / 145    | 0 / 124       |
| CFLGYFCTCYFGLFC  | 252 / 693     | 73 / 143   | 27 / 145   | 17 / 124      |
| FCFHKCFC | 4 / 693 | 5 / 143 | 1 / 145 | 0 / 124 |
| LVLWLLCNCFLC | 0 / 693 | 0 / 143 | 1 / 145 | 3 / 124 |

• Alignment frequency is presented as a ratio of the number of alignments that contain whole or part of the sequence with the highest percent identity to the total number of alignments.

Blastp search of CFLGYFCTCYFGLFC sequence reveals that it is part of non-structural protein 6 (nsp6) of replicase polyprotein 1a (ORF1a polyprotein) (*s6* of *s5–13*). Nsp6 is cleaved from ORF1a polyprotein. It is indicated at the UniProtKB (The UniProt Consortium, 2019) (P0DTC1) that the subcellular location of nsp6 is host membrane and it is a multi-pass membrane protein, as inferred from sequence similarity to ORF1ab (UniProtKB: P0C6X7). The position of CFLGYFCTCYFGLFC sequence is 3784–3798 region of ORF1a polyprotein, corresponding to 215–229 region of nsp6. It is a transmembrane region with helical topology.

*3.2 MHC class I-binding*

NetMHC 4.0, PickPocket 1.1, and NetMHCpan 4.1 tools are used to predict the binding of CFLGYFCTCYFGLFC to MHC class I proteins. It is indicated (Lundegaard et al., 2008) for an earlier version of NetMHC (v3.0) that the method was used to predict MHC-binding peptides in viral proteomes, including influenza, SARS, and HIV, with an average of 75%–80% confirmed binders of MHC.

Peptides of CFLGYFCTCYFGLFC have strong binding affinities for different HLA alleles, as predicted by distinct tools (Table 3, *s19*, *s20*). NetMHC predicted 7 SBs and by PickPocket predicted 4 SBs. In case of NetMHC, one 10mer is SB to HLA-B*15:01, and the rest are SBs to HLA-A*24:02. In case of PickPocket, one 9mer is SB to both HLA-A*02:01 and HLA-A*24:02, one 12mer is SB to HLA-A*02:01, and the remaining two are SBs to HLA-A*24:02. NetMHCpan did not predict any SBs (*s21*). In case of NetMHCcons (Table 3, *s22*), there are 5 SBs, in which one 12mer is SB to HLA-A*02:01, and the rest are SBs to HLA-A*24:02. As a result, 10mer YFCTCYFGLF and 11mer GYFCTCYFGLF are predicted as SBs to HLA-A*24:02, by NetMHC, PickPocket, and NetMHCcons. Accordingly, they are termed as SB-peptides in the rest of this work.

**Table 3.** Results of MHC class I binding predictions for peptides of CFLGYFCTCYFGLFC.• Those written bold are termed as SB-peptides in the rest of this work.

| MHC class I type | Length | **NetMHC (SB)** | **PickPocket (≥0.5)** | **NetMHCcons (SB)** | **NetCTLpan (E)** |
|---|---|---|---|---|---|

| HLA | mer | | | | | |
|---|---|---|---|---|---|---|
| HLA-A*24:02 | 8mer | | | | | GYFCTCYF |
| HLA-A*01:01 | 9mer | | | | | FLGYFCTCY |
| HLA-A*02:01 | | | YFCTCYFGL | | | |
| HLA-A*24:02 | | | | | | |
| HLA-B*15:01 | 10mer | FLGYFCTCYF | | | | |
| HLA-A*24:02 | | | | | | GYFCTCYFGL |
| **HLA-A*24:02** | | **YFCTCYFGLF** | **YFCTCYFGLF** | **YFCTCYFGLF** | **YFCTCYFGLF** | |
| HLA-A*24:02 | 11mer | | | | CFLGYFCTCYF | CFLGYFCTCYF |
| **HLA-A*24:02** | | **GYFCTCYFGLF** | **GYFCTCYFGLF** | **GYFCTCYFGLF** | **GYFCTCYFGLF** | |
| HLA-A*02:01 | 12mer | | FLGYFCTCYFGL | FLGYFCTCYFGL | | |
| HLA-A*24:02 | | LGYFCTCYFGLF | | | | |
| HLA-A*24:02 | | GYFCTCYFGLFC | | | | |
| HLA-A*24:02 | 13mer | FLGYFCTCYFGLF | | | | |
| HLA-A*24:02 | 14mer | CFLGYFCTCYFGLF | | CFLGYFCTCYFGLF | | |

⋅ NetMHCpan results are not displayed since there was no SB prediction there. (SB: strong binder; ≥0.5 stands for 1-log50k(aff)≥0.5; E: epitope.)

Prediction results for CFLGYFCTCYFGLFC are like that of nsp6 protein (*s23*) containing that 15mer (*s24–27*). SB-peptides are predicted as SB to HLA-A*24:02, also when the prediction of the 15mer is performed using the nsp6 protein.

All epitopes, except for one, are predicted to be binding to the HLA-A*24:02 allele (NetCTLpan results in Table 3, *s28*). One 9mer epitope is predicted to be binding to the HLA-A*01:01 allele. YFCTCYFGLF and GYFCTCYFGLF are predicted as epitopes of HLA-A*24:02, in accordance with the other predictions (Table 3). Those two epitopes are still predicted, and both are binding to HLA-A*24:02 when the 15mer target sequence is predicted as part of the nsp6 protein (*s29*, *note that the sequence is at the 214–228 position of the file*). C-term cleavage values are varying, as expected. Accordingly, SB-peptides YFCTCYFGLF and GYFCTCYFGLF of the 15mer are used for further analysis. These two SB-peptides are 91% overlapping, in other words, the 10mer is fully overlapped by the 11mer. Three-dimensional structure model of SB-peptide YFCTCYFGLF in complex with TCR alpha and beta chains, and MHC, is shown in Fig. 3.

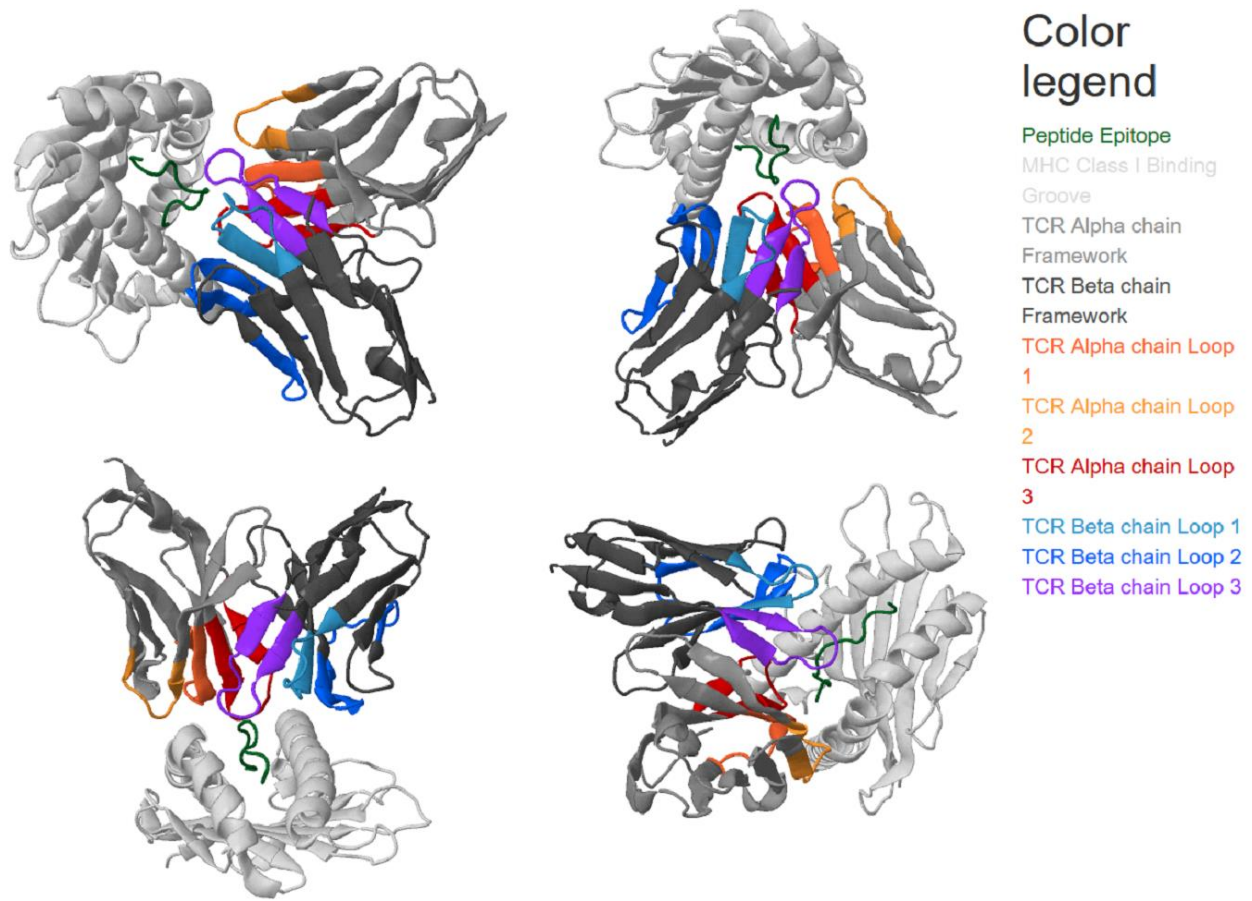

**Fig. 3.** Three-dimensional structure model of TCR alpha, TCR beta, YFCTCYFGLF, and MHC complex, from different perspectives. The model is obtained through TCRpMHCmodels (Jensen et al., 2019).

*3.3 Blastp search for SB-peptide and HLA-A*24:02 binding prediction of SB-peptide homologues*

*3.3.1 Blastp search and identifying SB-peptides' homologues*

Blastp search of CFLGYFCTCYFGLFC is performed by limiting search to *H. sapiens*. Top results in description table belong to immunoglobulin heavy chain junction region (*s30*). Searching for gapless alignments of two SB-peptides YFCTCYFGLF and GYFCTCYFGLF within results revealed no outcome for the latter. In case of the former, 5 sequences (YFCARNFGPF, YFCASSFGSF, YYCARRFGLF, YYCHCPFGVF, and YYCQQYFFLF) are identified (Table 4). The last one (YYCQQYFFLF) is not present when blastp search is set to return maximum 5000 target sequences instead of 20000.

Homologous sequences of SB-peptides are identified also in results of CFLGYFCTCYFGLFC blastp searches that are performed separately by limiting searches to individual Plasmodium species (*P. falciparum*, *P. malariae*, *P. vivax*, *P. ovale*, and *P. knowlesi*) (*s31–35*). Accordingly, there are 12 homologous sequences of the first SB-peptide, YFCTCYFGLF (Table 5, until the last row, *s36–40*). Search for homologous sequences of the second SB-peptide, GYFCTCYFGLF, yielded HYFPTITFHLF sequence of *P. ovale*,

*3.3.2 HLA-A\*24:02 binding predictions of SB-peptides' homologues*

After the identification of homologous sequences of SB-peptides, their binding to HLA-A\*24:02 is predicted. Accordingly, all the tools predicted YYCARRFGLF, YYCHCPFGVF, and YYCQQYFFLF sequences, to be binding to HLA-A\*24:02 (Table 4, *s41–45*).

**Table 4.** HLA-A*24:02 binding predictions for human 10mers homologous to the first SB-peptide, YFCTCYFGLF.

| Species | 10mer sequence | NetMHC (SB / WB / none)• | PickPocket (≥0.5 / <0.5)• | NetMHCpan (SB / WB / none)• | NetMHCcons (SB / WB / none)• | NetCTLpan (E)• |
|---|---|---|---|---|---|---|
| *H. sapiens* | YFCARNFGPF | SB | <0.5 (0.436) | WB | SB | E |
| | YFCASSFGSF | SB | <0.5 (0.456) | WB | SB | E |
| | **YYCARRFGLF** | **SB** | **≥0.5 (0.536)** | **SB** | **SB** | **E** |
| | **YYCHCPFGVF** | **SB** | **≥0.5 (0.554)** | **SB** | **SB** | **E** |
| | **YYCQQYFFLF** | **SB** | **≥0.5 (0.677)** | **SB** | **SB** | **E** |

• WB: weak binder; SB: strong binder; ≥0.5 and <0.5 stands for 1-log50k(aff)≥0.5 and 1-log50k(aff)<0.5, respectively; and E: epitope.

The sequences displayed in Table 4 belong to the following proteins in human proteome:

– YFCARNFGPF: Ig heavy chain variable region (Sequence ID (seq.ID) AAQ05358.1).

– YFCASSFGSF: T cell receptor beta chain variable region (seq.ID ANO56516.1).

– YYCARRFGLF: Immunoglobulin heavy chain variable region (seq.ID QGT38216.1).

– YYCHCPFGVF: Protocadherin FAT4 (seq.ID AHN13824.1), unnamed protein product (seq.IDs BAF84150.1, BAG53346.1), protocadherin Fat 4 isoform X2 (seq.ID XP_011530539.1), Fat tumor suppressor homolog 1 (seq.ID EAX05203.1), and protocadherin Fat 4 isoforms 3, 2, and 1 (seq.IDs NP_078858.4 and Q6V0I7.2, NP_001278214.1, and NP_001278232.1, respectively).

- YYCQQYFFLF: Anti-HIV immunoglobulin kappa chain variable region (seq.IDs AVQ94610.1 and AVQ94611.1).

As mentioned, there are 12 homologous sequences of YFCTCYFGLF, within Plasmodium species' proteomes (Table 5, until the last row, *s36–40*). Among those, 3 sequences are predicted as SBs to HLA-A*24:02 (Table 5, bold sequences until the last row, *s36–40*). Those are termed SB'-peptides. To determine if any of those SB'-peptides are homologous to human proteins while binding to HLA-A*24:02 with high affinity: First, blastp searches of SB'-peptides are performed separately, by limiting searches to *H. sapiens* (*s46–48*). One homologous sequence (YFYLFSLELF) is identified through analysis of blastp search results (Table 5, the last row). That is homologous to SB'-peptide FFYTFYFELF. It belongs to LOC441426 protein in human proteome (seq.IDs AAI36787.1 and AAI36791.1), and is also SB to HLA-A*24:02 (Table 5, the last row, and *s49–53*).

**Table 5.** HLA-A*24:02 binding prediction outcomes of peptides in Plasmodium species' proteomes that are homologous to the first SB-peptide YFCTCYFGLF (until the last row) and of the peptide in human proteome (the last row), which is homologous to the SB'-peptide FFYTFYFELF.·

| Species | 10mer sequence | NetMHC (SB / WB / none) | PickPocket (≥0.5 / <0.5) | NetMHCpan (SB / WB / none) | NetMHCcons (SB / WB / none) | NetCTLpan (E / none) |
|---|---|---|---|---|---|---|
| *P. falciparum* | FFCTPFFILF | WB | <0.5 (0.446) | WB | WB | E |
|  | YICIFYFILF | WB | <0.5 (0.345) | none | none | none |
|  | YIITCLSGLF | WB | <0.5 (0.288) | none | none | none |
| * | **FFYTFYFELF** | **SB** | **≥0.5 (0.500)** | WB | **SB** | E |
|  | YICTGMFSLF | WB | <0.5 (0.316) | none | WB | none |
| *P. malariae* | FFCTKYFAHF | WB | <0.5 (0.442) | WB | WB | E |
|  | **YFVACLFILF** | **SB** | **≥0.5 (0.503)** | WB | **SB** | E |
| *P. vivax* | YFATFYFTLY | SB | <0.5 (0.334) | none | WB | E |
| *P. ovale* | **YFPTITFHLF** | **SB** | **≥0.5 (0.589)** | **SB** | **SB** | E |
|  | YFGTFYFMLY | WB | <0.5 (0.347) | none | none | E |
| *P. knowlesi* | FFCACLFLLF | SB | <0.5 (0.472) | none | SB | E |
|  | YFATFYFTLY | SB | <0.5 (0.334) | none | WB | E |
| * *H. sapiens* | **YFYLFSLELF** | **SB** | <0.5 (0.455) | **SB** | **SB** | E |

· HLA-A*24:02 binding prediction results until the last row present predictions for Plasmodium species' sequences homologous to SB-peptide YFCTCYFGLF. Among those, the ones that are identified as SBs are written bold and termed as SB'-peptide. Results in the last row are the human peptide homologous to FFYTFYFELF, the only peptide identified at the end of the search for homologous peptides of SB'-peptides in human proteome. (WB: weak binder; SB: strong binder; ≥0.5 and <0.5 stands for 1-log50k(aff)≥0.5 and 1-log50k(aff)<0.5, respectively; and E: epitope.)

Peptide in Plasmodium species' proteomes homologous to the second SB-peptide GYFCTCYFGLF is the HYFPTITFHLF sequence of *P. ovale*. It is also predicted as SB to HLA-A*24:02 (*s54*). Analysis of blastp search results of that sequence in human proteome (*s55*) did not reveal homologous sequences.

*3.4 Visualization*

Fig. 4 displays sequence profiles of human- and Plasmodium-sourced 10mers that are both SBs to HLA-A*24:02 and have 6 identical residues with YFCTCYFGLF, compared to the 9mer motif of HLA-A*24:02. Three-dimensional structure model of TCR alpha, TCR beta, YFCTCYFGLF, and MHC complex is already presented (Fig.3).

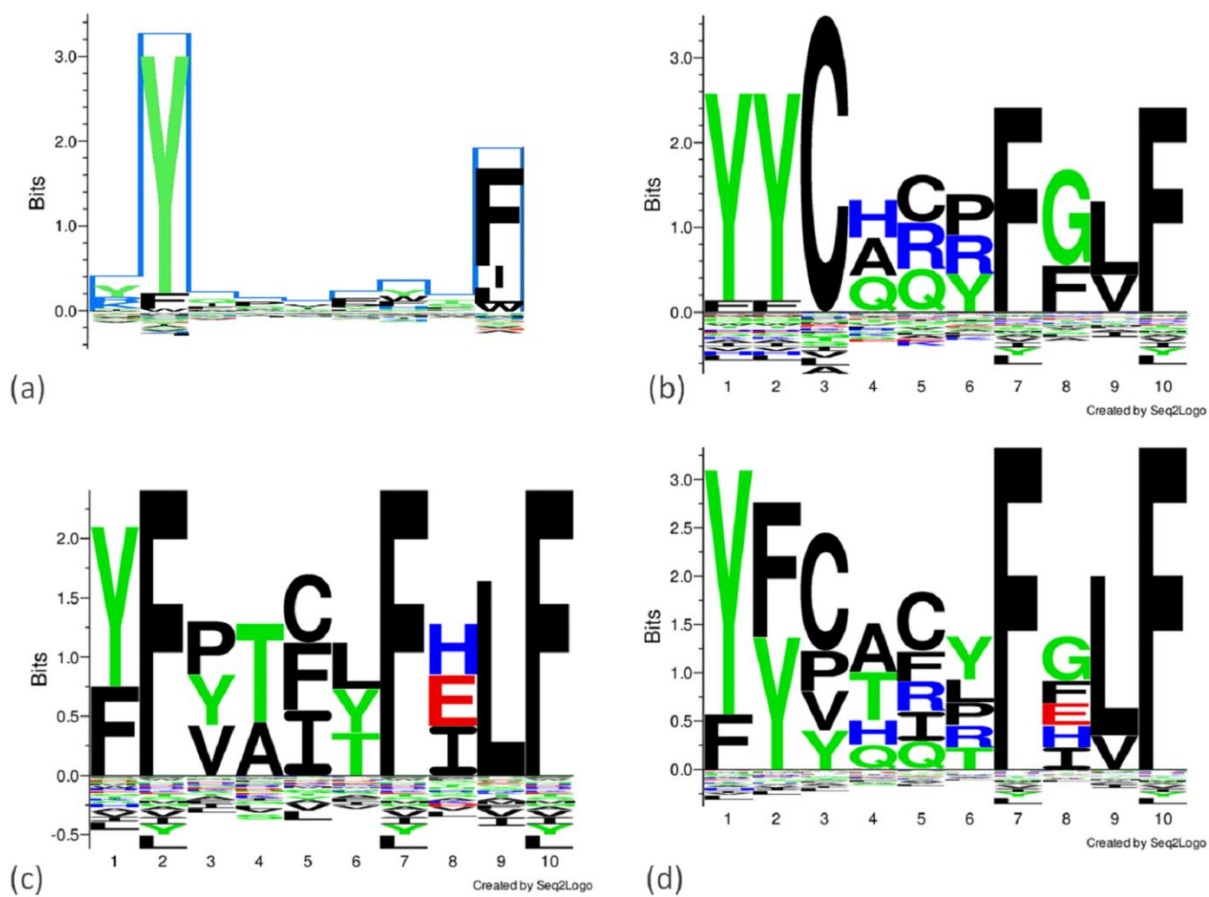

**Fig. 4.** Sequence logos of 10mers that are both SBs to HLA-A*24:02 and have 6 identical residues with SB-peptide YFCTCYFGLF, compared to HLA-A*24:02 motif. (a) is the 9mer motif of HLA-A*24:02. (b) is the logo generated from YYCARRFGLF, YYCHCPFGVF, and YYCQQYFFLF sequences (HLA-A*24:02 binding peptides in human proteome, homologous to YFCTCYFGLF). (c) is the logo generated from FFYTFYFELF, YFVACLFILF, and YFPTITFHLF sequences (HLA-A*24:02 binding peptides in Plasmodium species' proteomes, homologous to YFCTCYFGLF). (d) is the logo generated from sequences in (b) and (c). Dimensions of the original file of data in (a) is adjusted to comply with figure dimensions. Website of file for 9mer motif of HLA-A*24:02 in jpg format: https://services.healthtech.dtu.dk/services/MHCMotifViewer/HLA-A_files/Media/HLA-A2429/HLA-A2429.jpg?disposition=download.

## 4. Discussion

CFLGYFCTCYFGLFC sequence is identified here as query sequence with the highest identity to a to *P. vivax* sequence. This 15mer sequence is prevalently observed within tblastx alignment result and is common within respective results of the other Plasmodium species as well (Fig. 2, Table 2). It is suggested to be a transmembrane region (UniProtKB: P0DTC1). In relation, it is mentioned in other studies (Bianchi et al., 2017) that MHC class I molecules present transmembrane helices. YFCTCYFGLF part of this 15mer peptide is predicted as SB to HLA-A*24:02 of HLA supertype representatives. Yet, the other alleles can also be SB to parts of this 15mer and they are eliminated here because we looked for agreement of prediction results of different tools. YYCARRFGLF, YYCHCPFGVF, and YYCQQYFFLF sequences in human proteome, which are among homologous sequences of YFCTCYFGLF, are predicted as SB to the same HLA allele (Table 4, bold sequences). Upon being infected with SARS-CoV-2, individuals with HLA-A*24:02 can develop autoimmune responses, accordingly. Further, among sequences in Plasmodium species' proteomes homologous to YFCTCYFGLF, 3 sequences are predicted as SBs to the same HLA allele (Table 5, bold sequences until the last row). Those pathogen sequences can trigger the immune response in HLA-A*24:02 serotypes, upon getting infected with respective Plasmodium species. Yet, the question here is whether it can lead to an autoimmune response as well, through homology to SARS-CoV-2 peptide. Therefore, we looked if any of those are homologous to human proteins that are also SB to the same HLA allele. YFYLFSLELF is identified accordingly (Table 5, the last row).

Aligned SARS-CoV-2 query sequences with gaps are eliminated in this study. Additionally, aligned subject sequences with gaps in regions other than the positions that align with the investigated sequence regions were also eliminated. This process diminished the number of sequences that were obtained. In support, relevant work (Adiguzel, 2021) reported a higher number of results by presenting alignments of the same 15mer with human proteome, unrestricted to alignments' gaps. Yet, results outlined here are obtained although restrictions implemented.

Other than the discussions above, this study is conducted using reference genomes, meaning that individual variations in indicated sequences can influence outcomes, but they were not evaluated. Further genetic, physiological, and environmental variations can also contribute. All can influence possible immune responses and autoimmune reactions, but they cannot be evaluated in studies with single, representative reference genomes.

Peptides with homologous regions and strong binding affinities to the same HLA allele possess autoimmune reaction risk (Zabriskie and Freimer, 1966; Kohm et al., 2003; Yuki, 2007; Lule et al., 2017; Negi et al., 2017; Trost et al., 2010; Vellozzi et al., 2014). Intense rise in the number of shared heptapeptides among bacteria and human proteomes is reported (Trost et al., 2010). Yet, antibody generating pathogen proteins are not homologous to human proteins, and *vice versa* (Amela et al., 2007). These indicate that mere resemblance is generally not the only reason for developing autoimmunity, which should involve more parameters, including environmental factors (Rojas et al., 2018). Distinguishing self from non-self is a probable contributor (Matzinger, 2002). To account for molecular mimicry leading to an autoimmune disease, in addition to homologous host and pathogen proteins, the presence of cross-reacting T-cells or antibodies, epidemiological link and autoimmune disease generation, and reproducibility in an animal model, are required. In accordance, autoimmune reactions were observed in animal models (e.g., (Fujinami et al., 1983; Fujinami and Oldstone, 1985)). However, the latter two criteria are still challenging and prone to concerns (Rojas et al., 2018). In any case, findings of *in silico* studies as this one need to be supported by further evidence, in terms of autoimmunity susceptibility risk. Considering these criticisms, this work presents a condition of autoimmunity susceptibility risk of certain HLA serotypes, upon being infected with SARS-CoV-2 and *P. falciparum*, through homology to the same SARS-CoV-2 peptide.

Implications of the current study also highlight the importance of accounting for genetic variations in vaccine development. Far more genetic variations than HLA serotypes exist. Cross-reactivity between adjuvant-vaccine and patient proteome can possess autoimmune reaction risk in genetically susceptible individuals (Kanduc, 2012). Therefore, homologies between adjuvant vaccines and human proteome need to be investigated (Kanduc and Shoenfeld, 2016; Kanduc and Shoenfeld, 2018; Kanduc and Shoenfeld, 2019; Kanduc and Shoenfeld, 2020a; Kanduc and Shoenfeld, 2020b; Kanduc and Shoenfeld, 2020c) by taking genetic variations into account. Individuals to be included in the respective clinical trials should be representative of allelic variation of HLA types. As suggested (Nguyen et al., 2020), the implementation of HLA typing into ongoing and upcoming clinical trials and COVID-19 tests would be extremely useful. Such considerations are of importance for tests on animals as well since autoimmune responses would not necessarily be indicative of the same situation in humans, and *vice versa* (Kanduc and Shoenfeld, 2020b).

In case of COVID-19, HLA class I and II alleles' prediction from transcriptome sequencing data of bronchoalveolar lavage fluid samples of 5 COVID-19 patients from China also led to the identification of HLA-A*24:02, and it was deemed significant (Warren and Birol, 2020), compared to the frequency of the allele in South Han Chinese population (Middleton et al., 2003). Authors mentioned that HLA-A*24:02 (A24 allele group) was unknown as a risk factor for SARS, but it is associated with diabetes (Adamashvili et al., 1997; Noble et al., 2002; Nakanishi and Inoko, 2006; Kronenberg et al., 2012), which is known as a risk factor in COVID-19 (Guan et al., 2020). HLA-A*24 is reported as "an independent predictor of 5-year progression to diabetes in autoantibody-positive first-degree relatives" of patients with Type I diabetes (Mbunwe et al., 2013). Malaria is also associated with diabetes (Danquah et al., 2010; Acquah, 2019; Carrillo-Larco et al., 2019). In the mentioned COVID-19 study (Warren and Birol, 2020), HLA class II alleles DPA1*02:02 and DPB1*05:01 were found to be frequent as well. Those alleles are common in Han Chinese and associated with autoimmune diseases Graves' (Chu et al., 2018) and narcolepsy (Ollila et al., 2015). The latter one, DPB1*05:01, is associated with chronic hepatitis B in Asians, in addition to being a risk factor effecting viral infection clearance ability ((Kamatani et al., 2009; Ollila et al., 2015) *cited in* (Middleton et al., 2003)). Accordingly, class II allele-binding predictions of SARS-CoV-2 15mer identified here is planned as a future work.

*Evolutionary perspective*

Peptide sharing events or high number of matches in peptides potentially carry autoimmune response risk in individuals with genetic susceptibilities. Observation of matches in peptides of different species that possess such risks is expected. The probability of a hexapeptide to occur is 1/64000000 (Kanduc, 2020), which is calculated through equal likelihood of occurrence of 20 aa at any of the 6 positions in a hexapeptide. Proteome size of the organism, to which blastp search is limited, is the site to realise that possibility: 1/64000000. An unbiased proteome with 64 million aa would be sufficient for observing any 6mer for once, considering that such a proteome would be comprised of 64 million hexamers (*ignoring the dissection of proteome into proteins*). 122962 proteins with the total number of 80769298 aa are present in human proteome (retrieved from the protein-list at NCBI, genome assembly id 582967). This size of a proteome can let every possible 6mer to be observed once, at least. In this study, we looked for a minimum of 6 aa-matches in a 10mer and did not observe more than 6 aa-matches. The expected occurrence of 6 aa-matches in a 10mer is 210 and that of observing 6mers is 5. 210 is the number of 6-membered subsets in a 10-membered set, calculated by 10!/(6!x4!) (*each member is the aa at each position of the 10mer*). However, we observed only few sequences with 6 aa-matches (Table 4), namely, few of the 6-membered subsets. So, the observation is less than expected. This lessening is because the calculated number is expected number of observations for unbiased conditions, but human proteome is biased, and most proteins have several isomers. These results somehow support views about involvement of evolutionary processes between human hosts and SARS-CoV-2 (Kanduc, 2019; Kanduc, 2020), in relation to a different part of the concept. The evolution of viral proteome can be providing a selection advantage based on infectivity, or elimination by the host. For instance, proteome of *P. falciparum* strain 3D7 reference genome has 5387 proteins with total number of 4103467 aa (data retrieved from the protein-list at NCBI, genome assembly id 895506). It is about 20 times smaller than that of the human proteome. Yet, that proteome also revealed 5 peptides with 6 aa-matches to the 10mer query peptide, the first SB-peptide (Table 5). This amount can be interpreted as less bias in *P. falciparum* proteome than in human proteome. *P. falciparum* proteins do not have several isomers like in case of human proteins, which supports this interpretation. Still, this observation should be validated with a larger and varied query set. Currently, it is interesting that only one of the three SB'-peptides of Plasmodium species in Table 5 has a 6 aa-matching peptide in human proteome. This observation can be related to host-pathogen evolution or co-evolution of pathogens, or both. Also, 32 Plasmodium sequences are present among 17657 sequences in descriptions table of blastn search of SARS-CoV-2 with somewhat similar sequences (*s56*), which contributed to this view. In relation, 10mer part of SARS-CoV-2 peptide investigated here have homologous peptides with 6 aa-matches both in human proteome and in Plasmodium species' proteomes. Yet only one of those Plasmodium species' peptides has one such peptide in human proteome. In relation, shared immunodominant regions between *P. falciparum* and SARS-CoV-2 were suggested to be explaining low incidence of COVID-19 within the malaria-endemic belt (Iesa et al., 2020). Here, autoimmune reaction risk is studied through such a similarity. Accordingly, low incidence in malaria-endemic belt can be due to the higher number of homologous peptides of Plasmodium species and SARS-CoV-2. It can cause more protection by triggering immune response than an autoimmune reaction. This claim is supported here by that only one of the three SB'-peptides of Plasmodium species have a homologous peptide in human proteome that is SB to the same allele (Table 5). Accordingly, probability of a protective effect due to shared peptides between pathogens is higher than their risk of causing autoimmune response. This difference is because only a subset, i.e., a smaller number, of those shared peptides are SB to the same HLA allele and have a homologous peptide in human proteome, which is also SB to the same HLA allele. It can be interpreted also as follows: Host-pathogen evolution and pathogen co-evolution at the "malaria-endemic belt" worked towards diminished autoimmune responses that involve molecular mimicry. It can be protective against severe cases of COVID-19 with autoimmune reactions involving associated mechanisms. Yet, this scenario does not rule out the potential of inherently related autoimmune pathologies through peptide homologies. Besides, such similarities are expected, although physiological outcomes are complex. Still, results of *in silico* studies like this one require experimental validation or clinical support, or both.

Eventually, studies like this one can pave grounds to certain HLA serotypes' recognition as risk groups and identification of novel alleles (Marsh et al., 2010; Cheranev et al., 2021; Robinson et al., 2020). It also highlights the importance of relevant considerations in vaccine studies and can contribute to studies involving commonalities between pathogens of human host, driven by host-pathogen evolution. As outlined nicely (Córdoba-Aguilar et al., 2021), what is as much important is the possibility of lessons that will be learned from the current pandemic to trigger an understanding leading to the development of disease prevention and environmental protection strategies for the well-being of not only humanity but also nature itself.

## 5. Conclusion

In this study, 10mer peptide at positions 5–14 of CFLGYFCTCYFGLFC sequence of SARS-CoV-2 is identified as a possible relationship between SARS-CoV-2 and *P. falciparum*. Homologous 10mers of that peptide in human proteome are predicted to have strong binding affinities to the HLA-A*24:02 allele. Plasmodium species other than *P. falciparum* have homologous 10mers of that peptide with high affinities to the same HLA allele. However, only the 10mer of *P. falciparum* has a homologous 10mer in human proteome, with strong affinity to that HLA allele. These results imply autoimmunity susceptibility of HLA-A*24:02 serotypes through molecular mimicry in SARS-CoV-2 and *P. falciparum* infections, due to homology to the same SARS-CoV-2 peptide. This result is yet to be supported by clinical and experimental findings.

**Acknowledgements:** Ecology and Evolutionary Biology Society of Turkey is acknowledged.

**Supporting information:** Supplementary files available at shorturl.at/bmzP6

**Availability of data and materials:** Data as supplementary files submitted to Database: Mendeley (Dataset1_ADIGUZEL) and available as supplementary files under the shortened URL (URL shortened at *https://www.shorturl.at/*) of the Google Drive link.

**Conflicts of interests:** There are no conflicts of interests to declare.

**Funding:** This research did not receive any specific grant from funding agencies in the public, commercial, or not-for-profit sectors.

**Authors' contributions:** YA performed conception and design of the work; acquisition, analysis, and interpretation of the data for the work; drafting and revising the work to the final format.

**Ethics approval and consent to participate:** Not applicable.

**Consent for publication:** Not applicable.